# Liquid metal contact as possible element for thermotunneling

Avto Tavkhelidze*, Zaza Taliashvili, Leri Tsakadze, Larissa Jangidze and

Nodari Ushveridze

*Tbilisi State University, 13 Chavchavadze ave. , 0179 Tbilisi, Georgia*

We investigated a possibility of application of liquid metal contacts for devices based on thermotunneling. Electric and thermal characteristics of low wetting contact Hg/Si, and high wetting contacts Hg/Cu were determined and compared. We got tunneling I-V characteristics for Hg/Si, while for Hg/Cu, I-V characteristics were ohmic. We explained tunneling I-V characteristics by presence of nanogap between the contact materials. Heat conductance of high wetting and low wetting contacts were compared, using calorimeter measurements. Heat conductance of high wetting contact was 3-4 times more than of low wetting contact. Both electric and thermal characteristics of liquid metal contact indicated that it could be used for thermotunneling devices. We solved liquid Cs in liquid Hg to reduce work function and make liquid metal more suitable for room temperature cooling. Work function as low as 2.6 eV was obtained.

*Corresponding author: E-mail: avtotav@geo.net.ge

# Introduction

In recent years, much work was done on using electron tunneling for cooling applications. First theoretical investigation of cooling by means of electron tunneling was done to deal with overheating in single electron transistors [1]. Inside metal/insulator/metal (MIM) tunnel junction, electron tunneling takes place through an insulator layer. Due to high thermal conductivity of ultra thin insulator layer, MIM junctions exhibited large heat backflow. It reduces cooling efficiency. Solution to the heat backflow problem, using multiple MIM tunnel junctions connected in series, was offered by Huffman [2]. Unfortunately, device of this type was not realized because of its technological complexity. In Ref. [3, 4], we offerred design of tunnel junctions of metal/vacuum/metal (MVM) type, having very low heat backflow. Such tunnel junctions could be used for efficient cooling. Theoretical investigation of MVM tunnel junctions [5, 6] had shown that cooling coefficient could be as high as 20-30%. Efficiency could be further increased by introduction of insulator coating of the collector electrode (MVIM tunnel junction) Ref. [7]. In last case, potential profile inside the MVIM junction change in the way that, tunneling probability of high energy electrons is increased and tunneling probability of low energy electrons is reduced. Another method of using a vacuum gap, utilizing emission from semiconductor resonant states was proposed in Ref. [8].

Most cooling applications require tunnel junctions with a large area - of the order of square centimeter and more. The electrodes for tunnel junctions should be flat within few Angstroms to allow fabrication of uniform vacuum nano gap. Available polishing methods allow fabrication of surfaces with local roughness of 0.2 nm. However gradual deviation in the surface relief over large distances is as high as 500 nm per centimeter. The local roughness (0.2 nm) is low enough to obtain local vacuum tunneling, but because of a gradual deviation in the surface relief, it becomes impossible to bring large areas of two electrodes (polished independently) close enough to each other. We tried to solve this problem using liquid metal as one of the electrodes [9]. We assume that, in some cases, there was no direct thermal and electric contact between the liquid metal and the solid surface, and electrons tunnel between the electrodes. Given design has following technical advantages. Liquid metal automatically repeated the shape of the base electrode. Base electrode geometry change due to mechanical stress and thermal extraction, is automatically adopted by liquid metal electrode. There is no need in precise regulation of interelectrode distance.

Additionally, solid surface could be covered with thin insulator layer, to obtain MVIM junction. This will exclude direct electric contact between the electrodes and modify potential profile for high efficiency at the same time. Therefore, the above described design is potentially useful for the large area tunnel junction fabrication.

We experimentally investigated electric and thermal characteristics of low wetting contact Hg/Si(100) and high wetting contact Hg/Cu in exactly same conditions. To reduce the work function of liquid metal, Cs was solved in it and low work function Hg+Cs contact was investigated along with poor Hg contact.

## Sample Preparation and Experiment

Round Si (100) substrates with resistance <0.002 Ohm cm, having diameter of 20 mm and thickness of 2 mm were used as solid base for contacts. Si substrate was chosen as base electrode because it had very flat and uniform surface. Heavily doped n type Si was used to get contact close to MVM type. To achieve good thermal and electric contact with Si substrate (for high wetting contact Hg/Cu), Ti/Ag thin films were deposited on Si in vacuum ($10^{-6}$ Torr.). To remove contamination from Hg, it was first boiled in dimethilformamide during 3 minutes and was washed in deionizer water during 5 minutes. Next Hg was solved in $HNO_3$ (30% solution) and was mixed during 1 hour. $HNO_3$ solution was renewed 2-3 times during mixing. Next Hg was washed in deionizer water during 30 minutes and was stored under water. Conventional cleaning procedure was used for Si substrate. Additionally, thick Cu disc (0.5 mm) was grown electrochemically on Si/Ti/Ag substrate to obtain high wetting contact between Cu and Hg. To exclude contamination of Hg from atmosphere, measurements were made in vacuum environment $10^{-3}$ Tor.. Contacts with similar geometry were placed in vacuum chamber Fig.1. Si substrates (2) were placed on the Cu flange (1) and were thermally encored to it using liquid metal (In+Ga). Thin wall Teflon delimiters (3) were placed on Si wafers to contact edges. Pots obtained in this way were filled with liquid metal of equal weights. Cu rods (4), thermally encored to small external baths were places inside the liquid metal. External baths were used as heat reservoirs. Volume depicted by dash line on Fig.1 was evacuated. Cr/Al differential thermocouples were installed on the Cu flange and placed inside the liquid metal pots. Two Tungsten probes (not shown on Fig.1) were placed in each liquid metal pot to do four probe electric measurements. Another two contact probes were attached to Cu flange. Electric heater was attached to the Cu flange. During recording of the thermal end electric characteristics, both contacts were in the same vacuum and thermal conditions. Temperatures were monitored using differential and single thermocouples. Temperatures were stabilized within 1.5-2 C during measurements.

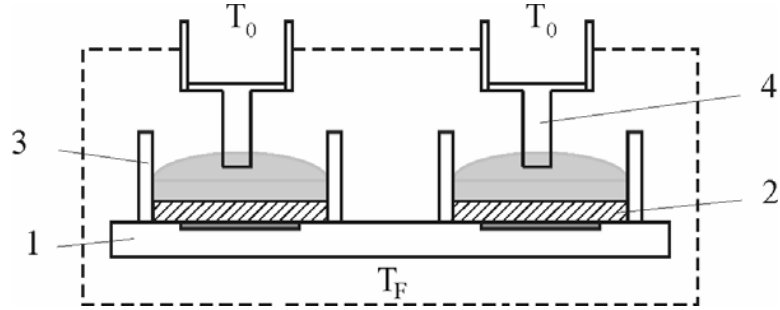

Fig.1. Experimental setup for measuring electric and thermal characteristics of wetting contacts.

To reduce work function of Hg (4.5 eV) we solved Cs in it. Cesiation of Hg was done by mixing CsCl crystals with liquid Hg and following heating to T=580 K during 10 hours. Work function of the mixture was measured using Kelvin probe. Minimum value of obtained work function was 2.6 eV which is in agreement with Ref. [10]. Work function value was stabile in time during days of measurements. All alkali materials solve in Hg and their solubility increase with temperature. Solubility of Cs was 6.5% at room temperature and 31% at 480 K Ref. [11]. To control solution we used X-ray analysis. Samples were cooled down in liquid nitrogen T=77 K prior to analysis, to transform sample to crystal form. We recorded peaks of Cs, Hg, CsCl and $Cs_2O$.

In order to obtain efficient cooling it was necessary that work function of both electrodes were reduced. So far, we obtained preliminary results for work function reduction in solid electrode [12, 13].

I-V characteristics of contacts were recorded using four point probe. In the case of low wetting I-V characteristics were tunneling and almost symmetric Fig.2b . Zero bias resistance varied between 3 and 9 ohms in separate experiments. In the case of high wetting I-V characteristics were almost ohmic (Fig.2a), with zero voltage resistance of the order of 15-18 m Ohm . Thus zero bias resistance ratio of high and low wetting contacts ($R_{Si}/R_{Cu}$) was in the range of $10^2$ - $10^3$. In the case of (Hg+Cs)/Si contacts I-V characteristics were tunneling and differed from I-V characteristics of Hg/Si contacts only quantitatively. Predominantly they had shown approximately 10 times more currents for the same voltages. However, tunneling current changed with time in irregular manner. Tunneling current differed dramatically from sample to sample. Instability could be explained by presence of CsCl grains in Hg+Cs mixture.

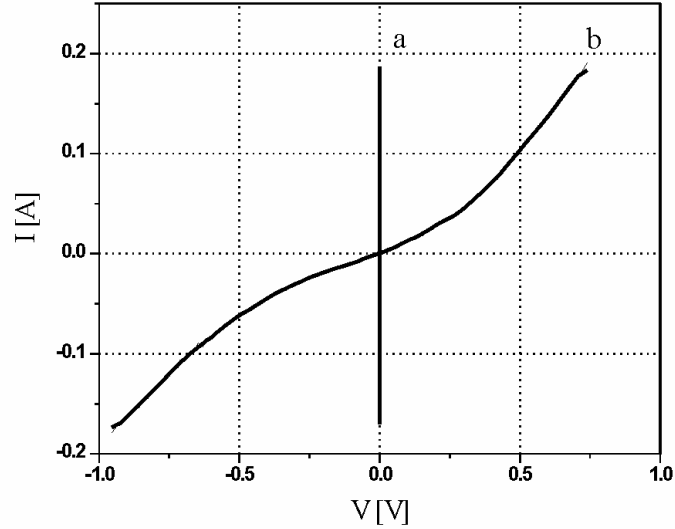

Fig.2. I-V characteristics of: (a) high wetting Hg/Cu contact and (b) low wetting Hg/Si contact.

To compare thermal conductance of low and high wetting contacts we applied temperature gradient between common Cu flange and thermal reservoirs and measured temperature gradients on contacts. Common Cu flange was heated using electric heater and thermal reservoirs were filled with water ice solution or liquid nitrogen. In our measurement setup the ration of thermal conductivities could be found using formula

$$\lambda_2/\lambda_1 = (T_{LM1} - T_0)(T_F - T_{LM2})/(T_{LM2} - T_0)(T_F - T_{LM1}). \qquad (1)$$

Here $\lambda_2$ and $\lambda_1$ are thermal conductivities of two contacts, $T_0$ is temperature of the external baths (Fig.1), $T_F$ is temperature of the common Cu flange, $T_{LM1}$ and $T_{LM2}$ are temperatures of the liquid metals in baths #1 and #2. Measurements were made for set of fixed temperatures and results were compared. For example, following set of temperatures $T_0$=273 K; $T_F$ =316.7 K, $T_{LM1}$=290.4 K and $T_{LM2}$=303.6 K give $\lambda_2/\lambda_1$ =0.28. Roughly same values were obtained for other sets of temperatures. Consequently thermal conductivity was 3-4 times higher for contact with high wetting Hg/Cu compared to contact with low wetting Hg/Si. Thermal radiation was not included in Eq.1. Radiation loses in our setup were negligible for temperatures we used.

We assumed that in the case of low wetting there was nano gap between the liquid metal and solid electrode. This assumption was based on following experimental results. First, I-V characteristics of low wetting contacts were tunneling unlike I-V characteristics of high wetting contacts which were ohmic. Second, zero point resistance of low wetting contact was much higher than of high wetting contact. Third, heat conductance of low wetting contact was considerably less than of high wetting contact. Our assumption was supported by modeling of liquid metals near low wetting surfaces. Modeling had shown that liquid creates intermediate layer of its vapor [14]. In order to verify our assumption we allowed Ar gas in vacuum chamber under pressure of 0.8 Bar. We did not found changes in I-V characteristics of low wetting contact introduced by Ar gas.

It should be noted that all types of contacts we investigated exhibit instability of electric characteristics in atmosphere. Even high wetting contact Hg/Cu, eventually had shown tunneling I-V characteristic. We explained it by absorption of air and water vapor on the surface. We think that it leads to creation of gas filled nanogap between the electrodes.

## Conclusions

Characteristics of low wetting Hg/Si and of high wetting Hg/Cu contacts were investigated in order to determine possibility of their application for thermotunneling. We found that low wetting Hg/Si contacts exhibit tunneling I-V characteristics and was potentially useful for thermotunneling. Zero point resistance of Hg/Si contact was $10^2$ - $10^3$ times more than of Hg/Cu contact. Thermal conductivity of low wetting Hg/Si contacts was 3-4 times less than of high wetting contacts Hg/Cu. Relatively low thermal conductivity of Hg/Si contact will allow its thermotunneling application. We made preliminary experiments to reduce work function of Hg by solving Cs in it. The work function, as low as 2.6 eV, was obtained. I-V characteristics of (Hg+Cs)/Si contact were tunneling as well. Due to reduced work function, tunneling currents were one order higher in (Hg+Cs)/Si contact.

## Acknowledgments

The work was supported by Borealis Technical Limited, assignee of corresponding US patents (7,253,549; 7,169,006; 7,140,102 ; 6,281,139; 6,417,060; 6,720,704; 6,774,003; 6,869,855; 6,876,123; 6,971,165).